\documentclass[prd,tightenlines,nofootinbib,showpacs,preprintnumbers,superscriptaddress]{revtex4}
\usepackage{amsfonts,amsmath,amssymb,amsthm,bbm,hyperref}
\usepackage{graphicx}
\usepackage{color}
\usepackage[T1]{fontenc}

\newcommand{\be}{\begin{equation}}
\newcommand{\ee}{\end{equation}}
\newcommand{\beq}{\begin{eqnarray}}
\newcommand{\eeq}{\end{eqnarray}}

\begin{document}

\title{Observational consequences of an interacting multiverse}
\author{Salvador J. Robles-P\'{e}rez}
\affiliation{Instituto de F\'{\i}sica Fundamental, Consejo Superior de Investigaciones Cient\'{\i}ficas, Serrano 121, 28006 Madrid, Spain,}
\affiliation{Estaci\'{o}n Ecol\'{o}gica de Biocosmolog\'{\i}a, Pedro de Alvarado, 14, 06411-Medell\'{\i}n, Spain.}
\date{\today}

\begin{abstract}
The observability of the multiverse is at the very root of its physical significance as a scientific proposal. In this conference we present, within the third quantization formalism, an interacting scheme between the wave functions of different universes and  analyze the effects of some particular values of the coupling function. One of the main consequences of the interaction between universes can be the appearance of a pre-inflationary stage in the evolution of the universes that might leave observable consequences in the properties of the CMB.
\end{abstract}

\maketitle

\section{Introduction}

The multiverse (see Ref. \cite{Carr2007} for a general review) has become the most general scenario in modern cosmology. However, the mayor controversy of the multiverse is its observability. The physical significance of the whole multiverse proposal roots on that feature. On the one hand, it could be thought that the multiverse is unobservable because whatever the definition of the universe is, it is always associated with some notion of causal closure. Thus, either all the events that are causally connected would belong to the same definition of the universe or, on the contrary, any event of other universe is not causally connected with the observable events of our universe and thus it cannot be observed. That is so, classically. Quantum mechanically, however, the quantum states of the matter fields that propagate in two distant regions of the whole spacetime manifold can be entangled to each other \cite{Mersini2008, Kanno2014} and, hence, their properties would be correlated. In that case, the boundary conditions to be applied on the scalar field should be such that they would consider the global state of the scalar field and not only particular states that correspond to each single region of the entangled pair separately.

The application of the boundary conditions to the global state of the scalar field and the lack of information about the state of the scalar field in the unobservable region makes  the quantum state of the scalar field in the observable region  be given by a reduced density matrix that contains the effects of the quantum correlations with the scalar field of the unobservable region. Thus, distant regions of the spacetime, which are not directly observable, can however leave some imprints in the properties of the matter fields of the observable region.

A similar reasoning can be made in the case of the wave function of the universe. Let us first notice that within the third quantization formalism \cite{Strominger1990, RP2010} the wave function of the universe can formally be seen as a scalar field that propagates in the minisuperspace of homogeneous and isotropic spacetimes and matter fields. In that case, the Wheeler-DeWitt equation can  formally be seen as the wave equation of a scalar field (the wave function of the universe), where the frequency of the wave equation is essentially given by the potential terms of the Wheeler-DeWitt equation. At the classical level, these terms are the potential terms of the Friedmann equation too. Therefore, the frequency of the wave equation that determines the evolution of the quantum state of the universe is ultimately related to the Friedmann equation that determines the classical evolution of the universes. An important feature is then that any interacting process that typically changes the frequency of the scalar field in a quantum field theory would change, in the parallel case of the multiverse, the potential term of the Wheeler-DeWitt equation and thus, it would have an observable consequence in the evolution of the universes.

In this paper, we shall pose an interacting scheme between the wave functions of different universes. As a result of the interactions the effective value of the potential becomes discretized. It turns out then that a new whole range of cosmological processes can now be posed in the multiverse, some of them would leave different imprints in the properties of the CMB \cite{Ade2016}. The paper is outlined as follows: in Sect. 2, we present the basics of the third quantization formalism, where an interaction scheme between universes can be posed in a similar way as it is done in the quantum mechanics of particle and fields. In particular, we explore in Sect. 2.3. the effects that different coupling functions might have in the evolutionary properties of the single universes. In Sect. 3, we briefly cite some ideas that have been proposed to test the multiverse. Afterwards, we explain how we expect to test the model of an interacting multiverse. We summarize and make some brief conclusions in Sect. 4.

%%%%%%%%%%%%%%%%%%%%%%%%%%%%%%%%%%%%%%%%%%
\section{Interacting multiverse}

%%%%%%%%%%%%%%%%%%%%%%%%%%%%%%%%%%%%%%%%%%
\subsection{Quantum multiverse}

Let us consider a homogeneous and isotropic spacetime with metric element given by
\be
ds^2 = - dt^2 + a^2(t) d\Omega_3^2 ,
\ee
where $a(t)$ is the scale factor and $d\Omega_3^2$ is the metric element on the three sphere of unit radius. Let us also consider a scalar field minimally coupled to the spacetime. The Hamiltonian constraint is then given by
\be\label{HC01}
\mathcal{H} \equiv -\frac{1}{a} p_a^2 + \frac{1}{a^3} p_\varphi^2 + \sigma^2 \left( H^2 a^4 - a^2 \right) = 0 ,
\ee
where \cite{Garay2014}, $\sigma^2 = \frac{3\pi M_P^2}{2}$ and $H^2(\varphi) = \frac{8 \pi}{3 M_P^2} V(\varphi)$. Following the canonical procedure of quantization the momenta are then promoted to operators and the Hamiltonian constraint (\ref{HC01}) transforms into the Wheeler-DeWitt equation, which can be written as \cite{Garay2014}
\be\label{WDW01}
\ddot{\phi} + \frac{1}{a} \dot{\phi} - \frac{1}{a^2} \phi'' + \omega^2(a,\varphi) \phi = 0 ,
\ee
where, $\phi \equiv \phi(a,\varphi)$, is the wave function of the universe \cite{Hartle1983}, the dot means derivative with respect to the scale factor and the prime denotes the derivative with respect to the scalar field. The frequency $\omega(a,\varphi)$ contains the potential terms of the Hamiltonian constraint (\ref{HC01}). It is given by
\be\label{FRQ01}
\omega^2 = \sigma^2 \left( H^2 a^4 - a^2 \right) . 
\ee
The Wheeler-DeWitt equation (\ref{WDW01}) has been written in a way that enhances the formal analogy with the wave equation of a scalar field. The scalar field to be quantized is now the wave function of the universe that propagates in the minisuperspace spanned by the variables, $q^N = \{ a, \varphi \}$, with metric element (of the minisuperspace) given by
\be\label{ME01}
d\mathfrak{s}^2 = G_{MN} dq^N dq^M = - a d a^2 + a^3 d\varphi^2 .
\ee
The third quantization formalism \cite{Strominger1990, RP2010} consist of considering and extending this formal analogy and quantize the wave function of the universe in a similar way as it is done in a quantum field theory. In particular, we can start by considering an action functional from which  the wave equation (\ref{WDW01}) can be obtained. It is given by\footnote{Let us add the superscript $3$ to the action, the Lagrangian and to the Hamiltonian density of the third quantization procedure.}
\be\label{3AC01}
^3S = \int da d\varphi \ \ ^3\mathcal{L}(\phi, \dot{\phi}, \phi'; a) ,
\ee
with
\be
^3\mathcal{L} = \frac{1}{2} \left( a \dot{\phi}^2 - \frac{1}{a} \phi'^2  - a \omega^2  \phi^2  \right).
\ee
Let us notice that in the metric element (\ref{ME01}) as well as in the action (\ref{3AC01}) the scale factor formally plays the role of the time like variable. Then, the momentum conjugated to the wave function of the universe, $\phi$, is given by
\be
P_\phi \equiv \frac{\delta (^3\mathcal{L})}{\delta \dot{\phi}} = a \dot{\phi} ,
\ee
and the Hamiltonian density is then 
\be\label{3H01}
^3\mathcal{H} = \frac{1}{2} \left( \frac{1}{a} P_\phi^2 + \frac{1}{a} \phi'^2 + a \omega^2 \phi^2 \right)  ,
\ee
which essentially is the Hamiltonian of a harmonic oscillator with \emph{time} dependent mass, $M(a) = a$, and frequency, $\omega(a,\varphi)$, given by (\ref{FRQ01}).

\subsection{Interacting scheme}

We can now pose an scheme of interaction between universes in a parallel way to how is done in quantum mechanics, by considering a total Hamiltonian given by \cite{RP2016}
\be\label{TH01}
\mathcal{H} = \sum_{n=1}^N \mathcal{H}_n^{0} + \mathcal{H}_n^I  , 
\ee
where $\mathcal{H}^{(0)}_n$ is given by (\ref{3H01}) for the universe $n$, and it corresponds to the Hamiltonian of a non-interacting universe. The interaction is described then by the Hamiltonian of interaction, $\mathcal{H}^I_n$. For this let us consider the following quadratic Hamiltonian
\be\label{HI01}
\mathcal{H}^I_n = \frac{a \lambda^2(a)}{8} \left( \phi_{n+1} - \phi_n \right)^2 ,
\ee
with the boundary condition,  $\phi_{N+1}\equiv \phi_1$. As it is well known in quantum mechanics, we can consider the following Fourier transformation
\be
\tilde{\phi}_k = \frac{1}{\sqrt{N}} \sum_n e^{-\frac{2 \pi i k n}{N}} \phi_n \ , \ \tilde{P}_k = \frac{1}{\sqrt{N}} \sum_n e^{\frac{2 \pi i k n}{N}} P_n ,
\ee
in terms of which  the normal modes the Hamiltonian (\ref{TH01}) turns out to represent $N$ non-interacting new universes, i.e.
\be
\mathcal{H} =  \sum \tilde{\mathcal{H}}_k^0 ,
\ee
where
\be
\mathcal{H}^0_k = \frac{1}{2} \left( \frac{1}{a} \tilde{P}_\phi^2 + \frac{1}{a} \tilde{\phi_k}'^2 + a \omega_k^2 \tilde{\phi}_k^2 \right) ,
\ee
with a new effective value of the frequency given by
\be
\omega_k^2(a,\varphi) = \sigma^2 \left( \tilde{H}_k^2 a^4 - a^2 \right) ,
\ee
with, $\tilde{H}_k^2 = \frac{8\pi}{3 M_P^2} \tilde{V}_k(a,\varphi)$, and
\be\label{MP01}
\tilde{V}_k(a,\varphi) = V(\varphi) + \frac{\lambda^2(a)}{4\pi^2 a^4} \sin^2\frac{\pi k}{N} .
\ee
The final result of the interaction is then an effective modification of the potential of the scalar field. However, it is worth noticing that the classical field equations are not modified because (here $\dot{\varphi}\equiv \frac{d \varphi}{d t}$)
\be
\ddot{\varphi} + \frac{3 \dot{a}}{a} \dot{\varphi} + \frac{d \tilde{V}_k}{d \varphi} = \ddot{\varphi} + \frac{3 \dot{a}}{a} \dot{\varphi} + \frac{d V}{d \varphi} = 0  .
\ee
The extra term in the potential (\ref{MP01}) entails a shift of the ground state that classically has no influence in the field equations. However, it determines the structure of the quantum vacuum states as well as the global structure of the whole spacetime.

\subsection{Modified properties}

Let us now consider several examples where the influence of the interaction among universes may be important. Let us first notice that the interaction among universes is not expected to have a significant influence in a large parent universe like ours. However, it may have a strong effect in small baby universes and thus, in the very early stage of the evolution of the universes. These effects may then propagate along the subsequent evolution of the universe and reach us as small corrections to the expected values of the non-interacting models of the universe (i.e. single universe models).

\begin{figure}
\centering
\includegraphics[width=12cm]{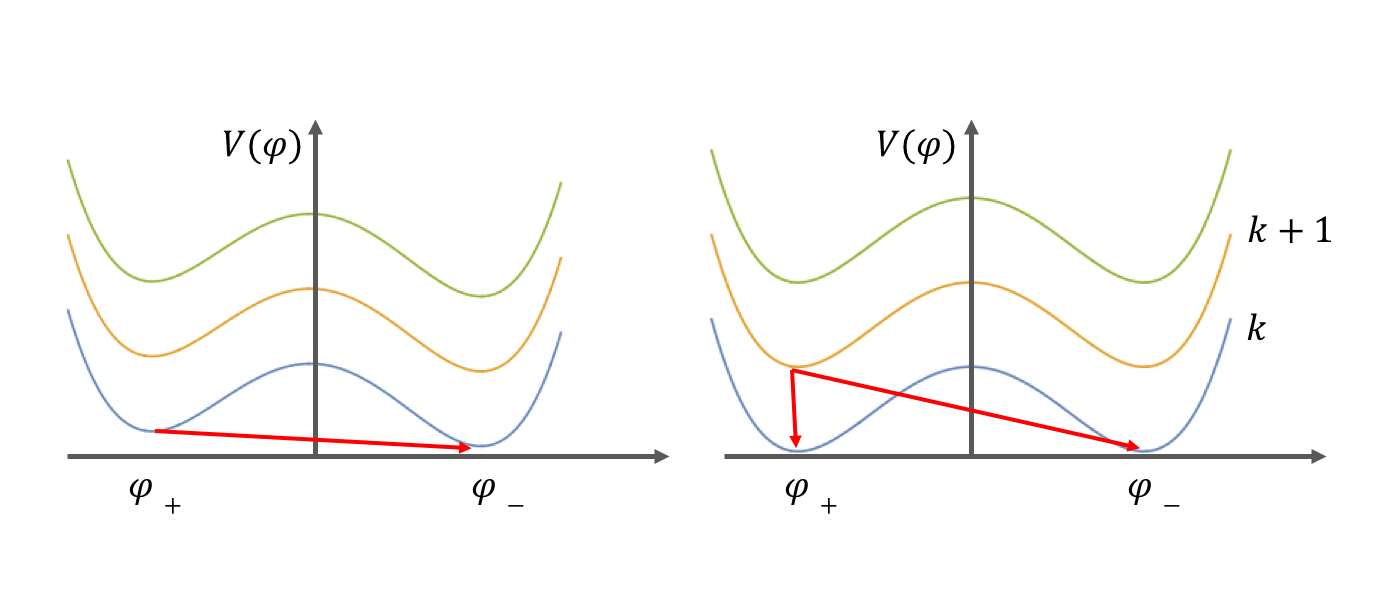}
\caption{The effect of the interaction among universe with a Coleman-DeLucia potential (left) and with a quartic potential (right). Ref. \cite{RP2016}}
\label{fig01}
\end{figure}

For concreteness, let us focus on the quartic potential studied in Ref. \cite{RP2016}. In that case, the interaction among universes induce a landscape structure of different false vacua and one or two true vacua (see, Fig. \ref{fig01}). Moreover, new processes of vacuum decay can now be posed including simple processes of vacuum decay as well as double decays that would lead to the formation of  entangled pairs of spacetime bubbles. The decaying rate per unit volume between two consecutive levels of the potential is given by
\be
\frac{\Gamma}{V} = A e^{-\frac{B}{\hbar}} ,
\ee
with \cite{RP2016}
\be\label{B01}
B = \frac{10 \pi^2 m^{12}}{3 \lambda_\varphi^8} \frac{1}{\left[ \sin^2\frac{\pi k}{N} - \sin^2\frac{\pi (k-1)}{N} \right]^3} \left( \frac{4 \pi^2 a^4}{\lambda^2} \right)^3 ,
\ee
where, $m$ is the mass of the scalar field, $\lambda_\varphi$, is the coupling of the quartic term in the potential, and $\lambda = \lambda(a)$ is the  coupling function that determines the interaction between the universes (see Eq. (\ref{HI01})). Eq (\ref{B01}) imposes a restriction on the values of the coupling function $\lambda$. Let us notice that in order for the vacuum decay to be suppressed  for large parent universes like ours, then, $\frac{\lambda^2}{a^4} \rightarrow 0 $, in the limit of large values of the scale factor. Even though, there are interesting cases fulfilling this condition with possibly observable imprints in the properties of the CMB of a universe like ours.

\begin{figure}
\centering
\includegraphics[width=7cm]{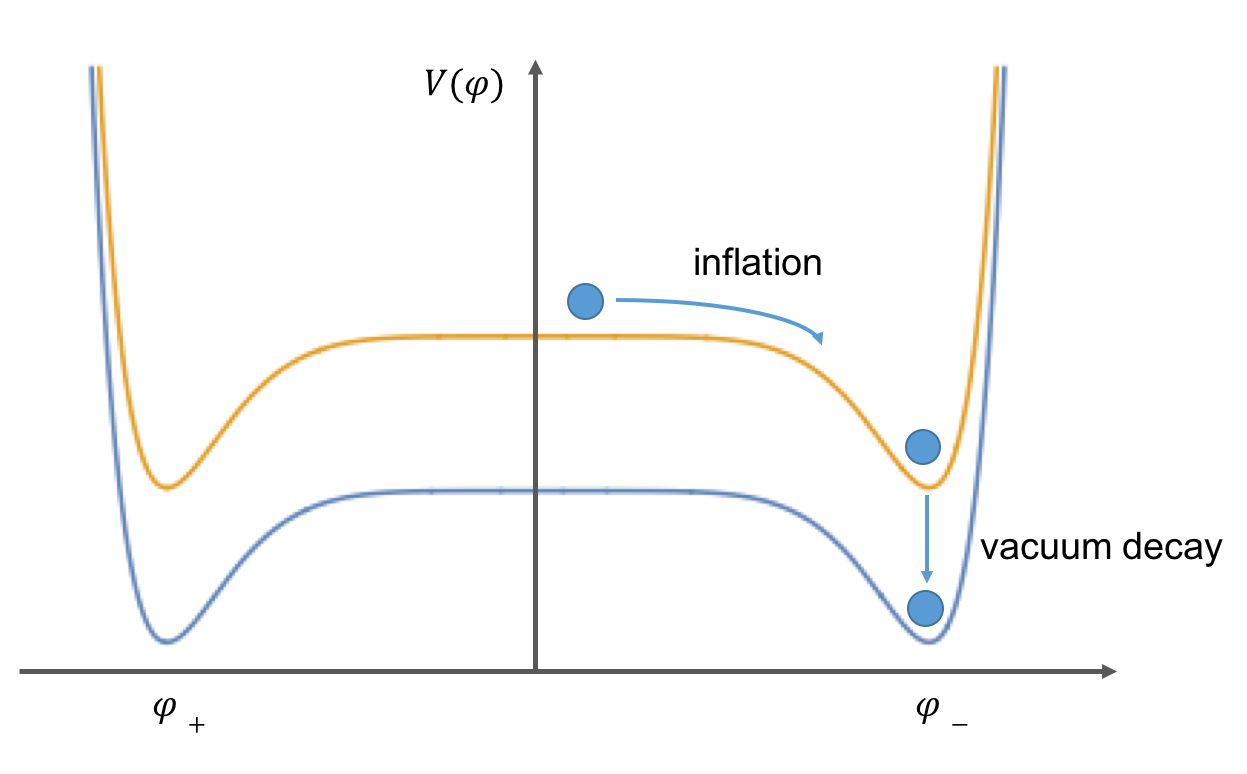}
\caption{Landscape structure for plateau like potentials. The potential energy of the plateau can be large enough to trigger inflation for a high value of the mode $n$. Afterwards, the universe may undergo a series of vacuum decays \cite{RP2016}.}
\label{fig02}
\end{figure}

For instance, let us consider the case where $\lambda(a) \propto a^2$. Then, during the slow-roll regime of the scalar field, for which $V(\varphi) \approx \Lambda_0$, the effective value of the cosmological constant turns out to be discretized as
\be
\Lambda^\text{eff}_k = \Lambda_0 + \Lambda \sin^2\frac{\pi k}{N} ,
\ee
with, $\Lambda^\text{eff}_k \in (\Lambda_0 , \Lambda_0 + \Lambda)$. If $\Lambda_0 \ll M_P^4$ and $\Lambda \sim M_P^4$, then, the interactions between universes might explain the apparent lack of potential energy for triggering inflation in the plateau like models that are enhanced by Planck \cite{Ijjas2013}. Despite the relative small value of the potential energy of the plateau with respect to the minima of the potential (see Fig, \ref{fig02}), the interactions among universes may excite the universe to a state of high value of $k$. There, the absolute value of the potential would be high enough to trigger inflation in the universe, which may  suffer afterwards a series of vacuum decays to reach a small value of the vacuum energy (see Fig. \ref{fig02}).

\begin{figure}
\centering
\includegraphics[width=7cm]{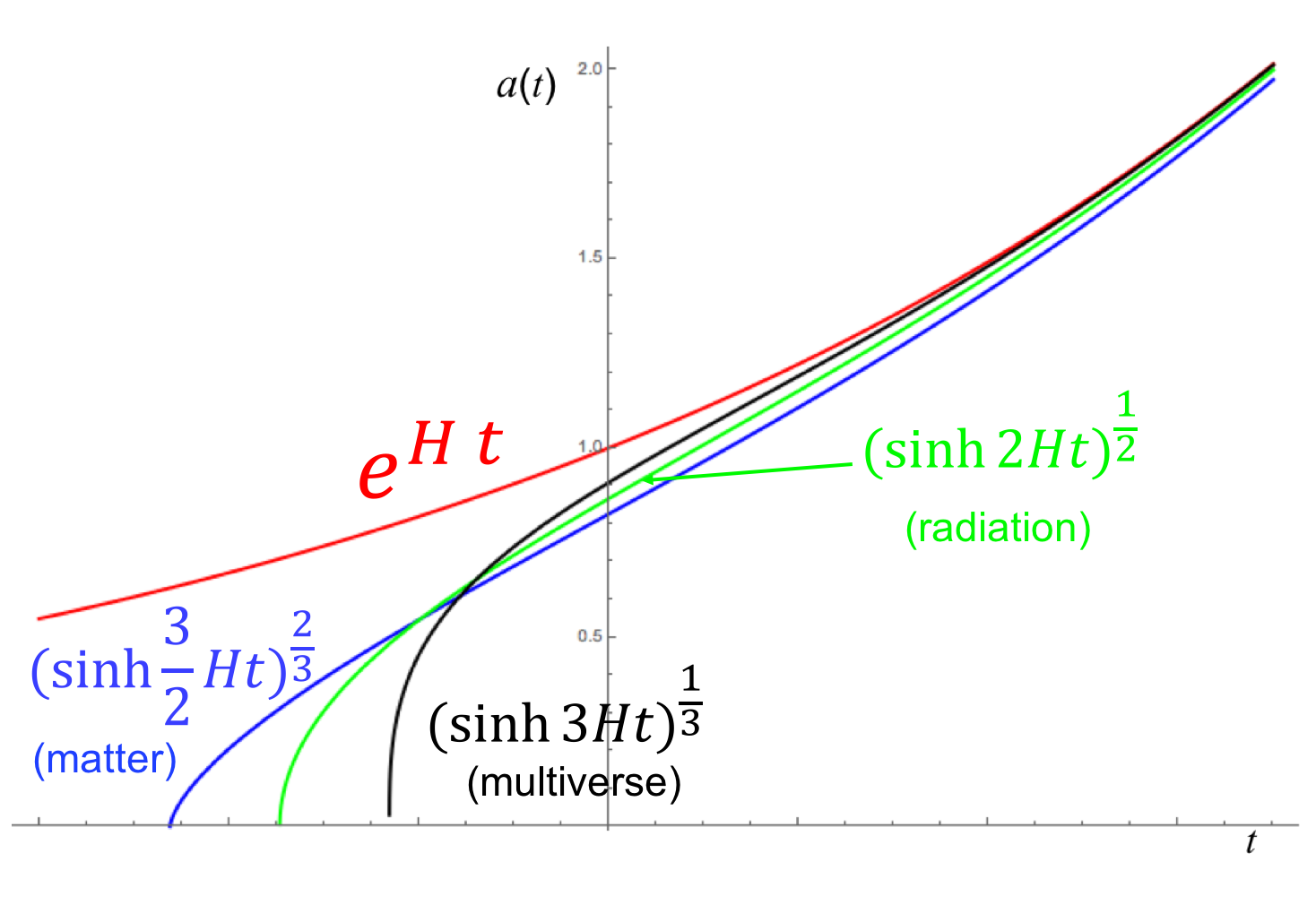}
\caption{The departure of different pre-inflationary stages from the exponential DeSitter evolution of the universe (red): matter-dominated evolution (blue), radiation-dominated evolution (green), and the evolution dominated by the quantum corrections induced by the interactions between universes (black). Ref. \cite{RP2016}}
\label{fig03}
\end{figure}

The second case that we can examine is the case where $\lambda$ is a constant. Then, the Wheeler-DeWitt equation in the $\tilde{\phi}_k$ representation  turns out to be \cite{RP2016}
\be
\ddot{\tilde{\phi}}_k + \frac{1}{a} \dot{\tilde{\phi}}_k - \frac{1}{a^2} \tilde{\phi}_k'' + \omega^2_k(a,\varphi) \tilde{\phi}_k = 0 ,
\ee
with
\be
\omega_k^2(a,\varphi) = \sigma^2(H^2 a^4 - a^2 + E_k) ,
\ee
and, $E_k = E_0 \sin^2\frac{\pi k}{N}$. It effectively represents the quantum state of a universe with a radiation like content. Let us notice that in terms of the frequency $\omega$, the effective value of the Friedmann equation is
\be\label{FR01}
\left(\frac{\dot{a}}{a}\right)^2 = \frac{\omega^2 }{a^4} \propto H^2 - \frac{1}{ a} + \frac{E_k}{a^4}  ,
\ee
where the last term is equivalent to a radiation like content for which the energy density goes like $a^{-4}$. For the flat branch the Friedmann equation (\ref{FR01}) can analytically be solved yielding
\be\label{SF01}
a(t) = a_0 \sinh^\frac{1}{2}(2H t + \theta_0) ,
\ee
where, $a_0$ and $\theta_0$, are two constants of integration. The scale factor (\ref{SF01}) departures from the exponential expansion of a flat DeSitter spacetime at early times (see, Fig. \ref{fig03}). This is a relevant feature because a radiation dominated pre-inflationary state in the evolution of the universe might have observable consequences in the properties of the CMB \cite{Scardigli2011} provided that inflation does not last for too long. It is  remarkable then that some interacting processes in the multiverse might have observable consequences in the properties of a single universe like ours, although in this case it would not be distinguishable from an ordinary radiation content of the early universe.

Let us finally consider the case where the coupling function is proportional to $a^{-1}$. In that case the frequency is given, in the limit of a large number of universes $N$, by
\be\label{FQ02}
\omega_k^2(a,\varphi) = \sigma^2 \left( H^2 a^4 - a^2 + \frac{ k^2 }{a^2} \right) .
\ee
The last term of the frequency (\ref{FQ02}) appears as well as a quantum correction to the Wheeler-DeWitt equation caused by the vacuum fluctuations of the wave function of the universe \cite{Garay2014}. It can be considered thus a sharp quantum effect having no classical analogue and thus a distinguishable effects of the interacting multiverse. The pre-inflationary stage induced in the evolution of the universe is more abrupt (a term $a^{-6}$ in the Friedmann equation) than those induced by a matter ($a^{-3}$) or a radiation ($a^{-4}$) content in the early universe. For the flat branch,
\be
a(t) = a_0 \sinh^\frac{1}{3}(3Ht + \theta_0) ,
\ee 
whose departure fro the exponential expansion of the DeSitter case is stronger than the other cases (see, Fig. \ref{fig04}).

\section{Observational imprints}

It is remarkable that we can now pose observable imprints of the multiverse in the properties of a universe like ours. This was probably unthinkable not so many years ago\footnote{To our knowledge, this was first done in Ref. \cite{Mersini2007, Mersini2008}. Before that, the multiverse was considered by many physicists to be an exotic proposal deprived of any scientific validation. Still today, there are many scientists that are sceptic about the whole proposal. However, we can now pose new and more precise observable imprints, and more and more groups are working on the search of the imprints of the multiverse \cite{Mersini2008, Kanno2014, Kanno2015, Bousso2014, Bousso2015, Garriga2016, Mersini2017, Divalentino2017a, Divalentino2017b, RP2017b}.}. There are several imprints that the existence of a multiverse would leave in the properties of a universe like ours, which can be classified in terms of the underlying phenomena. For instance, some authors have analysed  the effects that the entanglement between the modes of a scalar field that propagates along two causally disconnected regions of the spacetime would have in the spectrum of fluctuations \cite{Mersini2008, Kanno2014, Kanno2015}. The highest modes of the spectrum are unaware of the entanglement. However, for the wave lengths of order of the Hubble length the effect might be significant \cite{Mersini2008, Kanno2014}.

One can also analyse the effects that the entanglement between the corresponding wave functions of the universe (i.e. the wave function of the spacetime and matter fields) would produce in the effective value of the Friedmann equation \cite{Mersini2008, Garay2014, RP2017a}. This would also entail effects like an extra temperature dipole of the CMB, a suppression of the lowest modes of the power spectrum of the CMB or a suppression of the rms amplitude $\sigma_8$ \cite{Mersini2008}.

Another phenomena that would leave an observable imprint in the properties of our universe is the vacuum decay in the context of the multiverse. The sudden transition from a different vacuum state preceding the slow roll phase of the inflaton field would produce a suppression of the lowest modes of the power spectra that is also compatible with the observed data \cite{Bousso2014, Bousso2015}. Finally, one can also consider the spectrum of masses for the black holes originated in the context of the multiverse \cite{Garriga2016}. If the predicted spectrum would fit with observation it could be regarded as evidence for inflation and for the existence of the multiverse \cite{Garriga2016}.

\begin{figure}
\centering
\includegraphics[width=8cm]{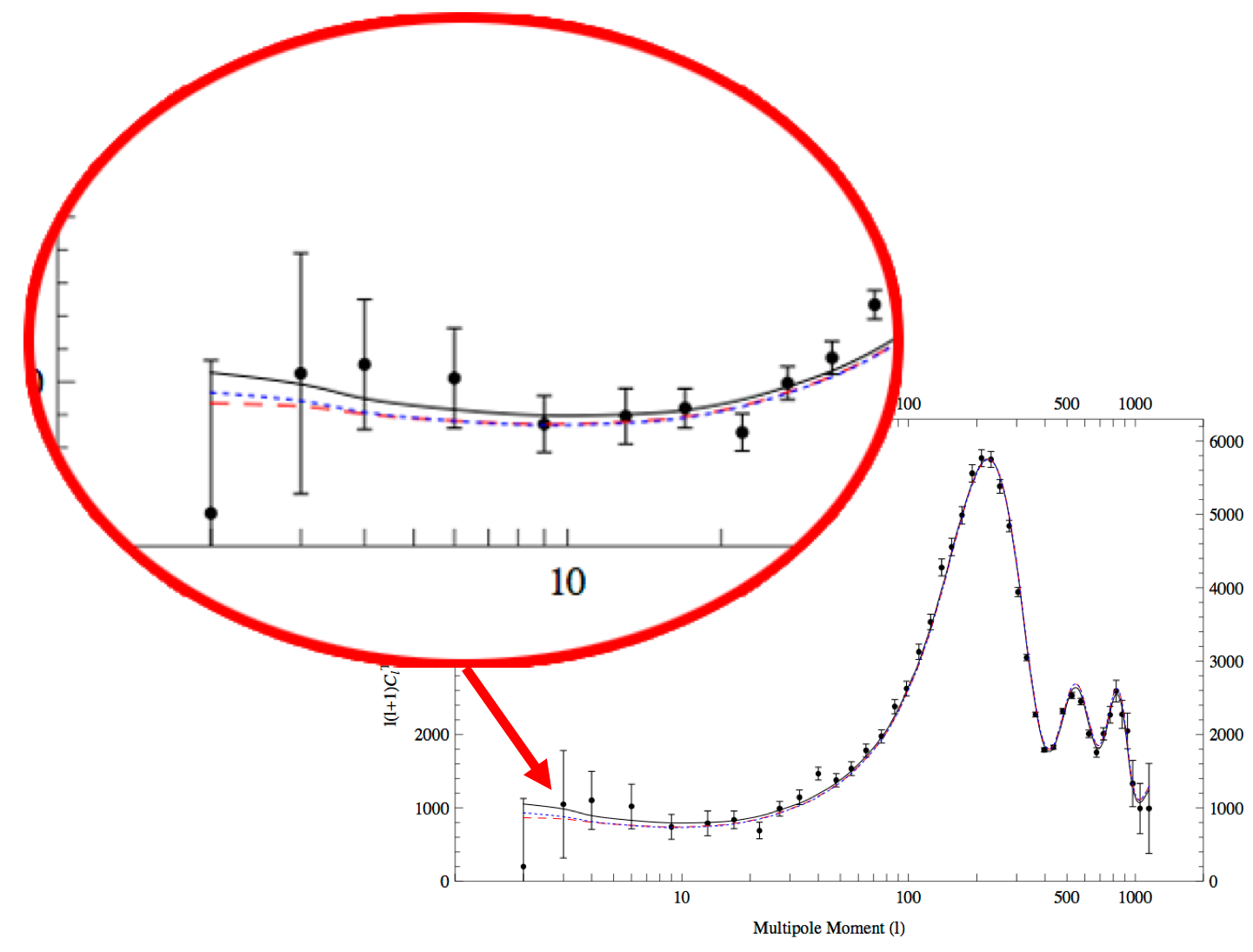}
\caption{Suppression of the lowest modes of the power spectrum of the CMB induced by the existence of a pre-inflationary stage of the universe dominated by: i) a network of frustrated domain walls (dotted blue line), and ii) a network of frustrated cosmic strings (dashed red line). Picture elaborated from the original figures of Ref. \cite{Bouhmadi2013}.}
\label{fig04}
\end{figure}

In this conference, we have shown the effects that an explicit interaction scheme among the universes of the multiverse would have in the effective value of the Friedmann equation. The interaction may first induce a discrete set of time dependent zero point values of the potential energy of the scalar field. It would have no influence in the classical equations of the scalar field. However, the quantum  states and the quantum fluctuations  would be affected by this new zero point energy. Furthermore, the interactions would entail new processes of vacuum decay like the creation of a pair of bubbles whose quantum mechanical states would be entangled \cite{RP2014}. In each single bubble, the interaction among universes would induce a pre-inflationary stage in the evolution of the universe that might have observable imprints, too. A pre-inflationary stage of the universe stem from the existence of different energy contents in the early universe has already been studied in the context of a single universe in Refs. \cite{Scardigli2011, Bouhmadi2011, Bouhmadi2013}, where the authors find that it can induce a suppression of the lowest modes of the power spectrum of the CMB that is compatible with the observed data (see, for instance, Fig. \ref{fig04}). However, the results are not conclusive, among different things because the error bars in the region of lowest modes of the CMB are too large to discriminate between different proposals. In principle, it  seems that the observational fit is better for a radiation like term ($a^{-4}$) than for a matter like term ($a^{-3}$), and that a stronger effect might even be needed to produce a better fit with observations \cite{Scardigli2011}. It is worth noticing that the term induced by the interacting multiverse ($a^{-6}$) is expected to produce a greater suppression of the lowest modes and thus a better fit \cite{RP2017b}. However, a more distinguishable prediction would be given by the fit with the observed data in the $l\sim 40$ peak of the CMB.

Within the interacting model of the multiverse presented in this talk, the next steps are clear. First, one can compute the effects that the pre-inflationary stage induced by the interactions between universes would cause in the power spectra of the CMB \cite{RP2017b}, and compare it with the latest astronomical data \cite{Ade2016}. One can also study the effects of the interacting multiverse in the properties of our universe by combining the procedures used in Refs. \cite{Bousso2014} and \cite{RP2016}. It is worth noticing that in the analogy between the third quantization formalism and the formalism of a quantum field theory presented in this talk, any interacting process that typically changes the frequency of the wave equation in a quantum field theory would modify in the parallel case of the multiverse the potential term of the Wheeler-DeWitt equation and, thus, it would generally have an influence in the effective value of the Friedmann equation through the relation (\ref{FR01}). Therefore, the kind of effects and interacting terms that can be analysed are quite large. The origin of the interacting terms would  be the low energy limit of the underlying theory, whether this is a string theory or the quantum theory of gravity. Therefore, the interacting multiverse could ultimately  be used to test these fundamental theories of modern cosmology, too.

%%%%%%%%%%%%%%%%%%%%%%%%%%%%%%%%%%%%%%%%%%
\section{Conclusions}

We have shown that the interaction among universes of the multiverse may modify the global properties of the single universes without changing their notion of causal closure. The effect of these interactions is expected to be  significant in the very early stages of the evolution of the universe where quantum corrections may be dominant. In that case, the interactions among the universe may create a landscape structure of different solutions among which the state of the universes can undergo different quantum transitions. For instance, it might well be that the universe would start in a quantum state of a high value of the mode, for which the absolute value of the potential is high enough to trigger inflation, and then it may suffer several process of vacuum decays until it would finally reach a state with a small value of the vacuum energy.

In general, any interacting process that might occur among universes of the multiverse would modify the effective value of the potential terms of the Wheeler-DeWitt equation, which are eventually connected with the Friedmann equation. Therefore, the interactions among universes of the multiverse modify the evolution of the universes. This is an effect that rapidly disappears as the universes expand. However, it may be significant on the earliest stage of their evolutions. In that case, they would leave observable consequences in the properties of the CMB, at least in principle, provided that inflation does not last for too long, i.e. in a scenario of \emph{just enough inflation}. Particularly, they would induce a suppression of the lowest modes of the power spectrum of the CMB that not only in compatible with the observed data but that it might even produce a better fit.

Probably the most important conclusion is that all these proposals for the testability of the multiverse bring it to the same footing of testability than any other cosmological theory. This is a crucial step in the formulation of the multiverse theories because the main controversy has  always been the apparent lack of testability of the whole proposal. The interacting multiverse not only makes is feasible but it would eventually serve to test the underlying theories of the string theories and the quantum theory of gravity.

\acknowledgments

This research was supported partially by the project FIS2012-38816, from the Spanish Ministerio de Econom\'\i a y Competitividad.

\bibliographystyle{apsrev}
%\bibliography{bibliography}

\end{document}